\documentstyle[12pt,aasms4]{article}
\input{psfig}
\eqsecnum
\def \m{\ifmmode M_\odot\else M$_\odot$\fi}
\def \r{\ifmmode R_\odot\else R$_\odot$\fi}

\def\kms{km~s$^{-1}$}

\def\beq{\begin{equation}}
\def\eeq{\end{equation}}
\def\ref{\reference}
\def\gr{$\gamma$-ray}
\def\grb{$\gamma$-ray burst}
\def\grbs{$\gamma$-ray bursts}
\def\ul{\underline}
\begin{document}
\title{The Supernova-Gamma Ray Burst Connection}
\author{Lifan Wang $^1$ and J. Craig Wheeler$^1$}
\affil{$^1$Astronomy Department, University of Texas, Austin, Texas 78712;
lifan@tao.as.utexas.edu, wheel@astro.as.utexas.edu} 

\begin{abstract}

Supernovae 1998bw and its corresponding 
relativistically expanding radio source are coincident with 
the \grb\ source GRB 980425. 
We show that of six recent SN Ib/c for which an 
outburst epoch can be estimated with some reliability, four have 
radio outbursts and all are correlated in time and space
with BATSE \grbs.  The joint probability
of all six correlations is 1.5$\times10^{-5}$.
No such correlation exists for SN Ia and SN II.  
The \gr\ energy associated with the SN/GRB events is
$\sim10^{46} - 10^{48}$ ergs if emitted isotropically.  Economy
of hypotheses leads us to propose that all \grbs\ 
are associated with supernovae and that the \grb\ 
events have a quasi-isotropic component that cannot be observed
at cosmological distances and a strongly collimated and 
Doppler-boosted component that can only be seen if 
looking nearly along the collimation axis.  
Such collimation requires a high rate of occurrence perhaps
consistent with a supernova rate. The collimated flow may be 
generated by core collapse to produce rotating, magnetized neutron stars.   
All core collapse events may produce such jets, but 
only the ones that occur in supernovae with small or missing 
hydrogen envelopes, Type Ib or Ic, can propagate into the interstellar 
medium and yield a visible \grb.  We suggest that
asymmetries in line profiles and spectropolarimetry
of SN II and SN Ib/c, pulsar runaway velocities, soft \gr\ 
repeaters and \grbs\ are associated phenomena. 

\end{abstract}


\section{Introduction}

The first few months of 1997 brought a major breakthrough in 
one of the outstanding mysteries of modern astrophysics, 
the cosmic \grbs\ (Klebesadel, Strong \& Olson 1973,  
Fishman 1995). The quest for an explanation 
of \grbs\ had been handicapped by a lack of direct 
knowledge of the distance to the bursts.  
This changed on February 28, 1997, when 
BeppoSAX localized a burst, GRB 970228, sufficiently well that 
an optical follow-up was feasible (Frontera et al. 1998).  The result was the 
discovery of the first optical counterpart (van Paradijs et al. 1997). 
Two months later, in early May, BeppoSAX found another \grb\ enabling 
another optical identification (GRB 970508; Bond 1997).  
In this case, foreground absorption lines show that the source is at
a redshift in excess of z = 0.863, proving
that the source is at a cosmological distance (Metzger et al. 1997).  
These two optical counterparts have revolutionized the field and proven
the power of focusing optical astronomy on this decades-old problem.   

Early estimates of the energy of \grbs\ assuming
cosmological distances gave energies comparable to the 
kinetic energy of supernovae, $\sim10^{51}$ ergs, entirely in
$\gamma$-rays.  From the lower limit to the redshift of the optical 
counter-part of GRB 970508 and the observed maximum 
brightness, $R\ \approx\ 19.6$ mag (Castro-Tirado et al. 1997), 
one can deduce a distance and hence the absolute brightness.  For
isotropic emission, this event was about a factor of 100 brighter 
than a Type Ia supernova at a redshift of 1.
A third optical counterpart was associated with GRB 971214.
The apparent host galaxy for this event has a redshift of 
z = 3.42 (Kulkarni et al. 1998a). For isotropic
emission, this event requires $3\times10^{53}$ ergs.

\section{SN 1998bw = GRB 980425}

Soffitta et al. (1998) reported the detection by the BeppoSAX 
Gamma-Ray Burst Monitor of a \grb\ on Apr. 25.90915 UT
(GRB 980425).  The BATSE experiment on CGRO confirmed the 
detection (Kippen et al. 1998).  The event was average in terms of the 
BATSE burst flux/fluence distribution.  
Galama et al. (1998) reported supernova 1998bw offset from the nucleus
of the face-on barred spiral galaxy ESO 184-G82.  This position is 
within the error box of GRB 980425, but does not coincide with either of 
the two X-ray sources of the BeppoSAX NFI image
(Pian, et al. 1998).  Lidman, et al. (1998) 
reported peculiar spectral properties for the supernovae.
Wieringa et al. (1998) reported no radio sources were detected within 
the 1' error radius of the two NFI sources, but that there is a bright 
radio source that coincides with the optical astrometric position 
given by Galama et al. (1998). 
At the distance to ESO 184-G82, the radio 
source was more than three times as 
luminous as SN 1988Z, one of the 
most luminous radio supernovae.  Given the seemingly small likelihood of 
finding the radio source and the supernova in this field, it is very 
probable that these events are associated with GRB 980425 despite the 
apparent discrepancy of the positions with the NRI locations. This
is supported by analysis of the radio source that strongly suggests 
that it is expanding relativistically (Chevalier 1998, Kulkarni et al. 1998b).  
Independent of the association with the $\gamma$-ray burst, this
is direct evidence that supernovae can produce relativistic outflows.

The association of a supernova with GRB 980425 may substantially alter 
the interpretation of the GRB afterglow.  The current fireball model 
involves the shock interaction between relativistic ejecta and the CSM/ISM.
The association with a supernova means that this interaction is {\it not} the 
only mechanism that produces optical emission although the interaction
is certainly important for producing the radio emission and perhaps the 
$\gamma$-rays and X-rays (M\'esz\'aros \& Rees 1997a).  
The UV/optical emission of SN 1998bw 
is likely to consist of two parts, one from the stellar 
ejecta and the other from synchrotron emission produced in accord with the 
popular fireball model. 

We have observed something similar to SN 1998bw before.
The optical spectra of SN 1998bw resemble those of SN 1997ef.  
SN 1997ef and SN 1998bw are the only two supernovae known for which 
high quality spectroscopy was obtained and yet they can not be easily
classified under the current classification scheme. At early times, 
SN 1997ef showed broad emission/absorption features with FWHM 
about 50,000 \kms, and all the features evolved rapidly 
to the red at a rate of about 1500 \kms day$^{-1}$
during the first week of discovery (Garnavich et al. 1997).  
At later times, the spectrum of SN1997ef 
resembled a Type Ic supernova which is characterized by little evidence 
for hydrogen, helium, or the strong Si II line of Type Ia.
Sadler et al. (1998) suggest an association of SN 1998bw with a
Type Ib, but it is clear that early spectra did not resemble
canonical SN Ib or SN Ic. Optical spectropolarimetry of SN 1997ef did not 
reveal polarization to an upper limit of 0.8\% (Wang, Howell, Wheeler 1998). 

\section{Other Supernova - Gamma-Ray Burst Associations}

We have investigated the correlation of supernovae with BATSE
$\gamma$-ray burst sources.  Table 1 shows a compilation of all
recent supernovae that had sufficiently well-established light curves
to define maximum light and hence to yield an estimate of 
the time of explosion.  Column 1 gives the identification of the supernova;
column 2, the spectral type; column 3, the estimated dates of 
the explosion using the same notation as for $\gamma$-ray bursts,
year, month, and day; column 4, the right ascension; 
column 5, the declination; column 6, whether there was a reported 
radio counterpart; column 7, the probability of at least one
random association with a $\gamma$-ray burst, given as
$1 - \prod_{i=1}^{N}(1-P_i)$, where $P_i$ is the BATSE 3-$\sigma$ error
box centered on the \grb\
divided by 4$\pi$ steradians for each $\gamma$-ray burst
that occurred in the supernova temporal window of column 3 and
N is the total number of these error boxes.
Where more than one number is
given in column 7, the probabilities were calculated using the 3-, 2-, and 
1-$\sigma$ BATSE error boxes, respectively. The underlined numbers indicate 
\grb\ counterparts were found within the error boxes
of the corresponding \grbs, whereas
non-underlined numbers indicate no associations. 
For the events where there is an associated 
$\gamma$-ray burst, the burst identification is given in column 1, 
and the RA and declination of the \grb\ in columns 4 and 5. 

Table 1 shows that of the 6 well-observed SN Ib/c since CGRO/BATSE
was launched, 4 had observed radio outbursts and {\it all} were
correlated with a \grb.  Four SN Ib/c were found to be within
the 2-$\sigma$ error boxes of the associated \grbs. 
The individual probabilities of random associations range from 0.06 to 0.3.  
The joint probability 
that all 6 associations are by chance is 1.5$\times10^{-5}$.  The real
probability of accidental association is probably much smaller
since the \grbs\ tend to occur at the best estimate of the time
of the supernova explosion.  The chance association in each case
with the error box of the particular \grb\ we have identified
is about $10^{-3}$ and the joint probability of the 6 events
is vanishingly small.  

Table 1 also gives data on 7 Type Ia events
that are expected to undergo thermonuclear explosion and hence
not core collapse.  There is {\it no} correlation of any
\grb\ with any of these events.  Table 1 also contains 3 Type II
events, none of which is correlated with \grbs\ and 4 events
like SN 1993J (referred to here as SN IIt).   These events are
thought to have only thin hydrogen envelopes and hence to be
nearly bare cores like SN Ib/c.  Of the SN IIt in Table 1,
there are no apparent correlations with \grbs\ except with
SN 1993J, the nearest and brightest.  A correlation with SN 1996cb is
excluded only by dint of the very small BATSE error box for
GRB 961212. 

To assess the probability of random association, we
have applied exactly the same test for each supernova, but artificially 
placed all the supernovae on the exact opposite side of the sky. Only three
supernovae, SN 1997ef (peculiar Ic), SN 199bq (Ia), and 
SN 1998T (Ib/c) were found to be within the 3-$\sigma$ error boxes of
3 \grbs\ that occurred at epochs corresponding to the temporal windows
for the supernova explosions. 
Considering that the probability for each association by random
is about 20\%, this number of correlations
is entirely consistent with chance in a sample of 21 supernovae. 

SN 1997ef was discovered about 10 days before maximum on JD 10778.  
As shown in Table 1, it was within the 2-$\sigma$ BATSE error box of 
GRB 971115 on JD 10768 (Garnavich, 1998) and within the 
3-$\sigma$ error box of GRB 971120 on JD 10772. At most one
association is real. Interestingly, SN Ib/c 1997ei 
was discovered at about the same time in 
roughly the same area of the sky.  SN 1997ei is found to be within
the 3-$\sigma$ error box of GRB 971120. 

\section{The Case for Strong Collimation}

The connection between \grbs\ and supernovae may call for 
a re-interpretation of the physical 
nature of $\gamma$-ray bursts.  
From the $\gamma$-ray fluence of GRB 980425 and the recession 
velocity (assuming $H_0$ = 65 \kms Mpc$^{-1}$)
the total \gr\ energy is $7\times10^{47}$ ergs. 
The \gr\ energy of the other relatively nearby 
events associated with SN Ib/c in Table 1 is $\sim10^{46}-10^{48}$ erg 
assuming spherical geometry. This is completely consistent with an
origin in supernovae.  The question arises as to the nature of
the very distant, apparently very energetic \grbs.  These
could also be related to supernovae if a portion of the 
\gr\ energy is strongly collimated.
In particular, the supernova/$\gamma$-ray 
burst connection suggests that we are dealing with energies that 
are a fraction of a supernova energy, perhaps ten percent, not 100 or 
more times a supernova energy. 
Strong collimation, in turn, demands a much higher 
incidence rate.  If, for example, about 10$^{50}$ ergs were collimated 
into 0.1 percent of $4\pi$ steradians in the form of a highly relativistic
jet, the equivalent energy would be 10$^{53}$ ergs if  
sphericity were assumed. The required rate of events would be 
correspondingly higher by a factor of 10$^3$, so \grbs\ 
would have to occur 
about once in 100 years per bright galaxy, not once per $10^5$ years, the 
canonical number for \grb\ and one associated 
with neutron star coalescence rates (Narayan, Piran \& Shemi 1991).
This rate could be lower by more than two orders of magnitude if 
the redshift for \grbs\ is as high as z $\sim$ 6, giving a
larger volume and lower rate per bright galaxy.
A frequency of once per 100 years is a conventional rate for supernovae.

\section{Discussion}

Table 1 strongly suggests that 
at least one component of the \grbs\ arise in supernovae. 
The data shows that normal SN II and SN Ia are unlikely to produce \grbs.
Table 1 does suggest that SN Ib/c are important \grb\ producers.
Woosely (1993; see also Fryer \& Woosely 1998)
anticipated the utility of bare stellar cores as a site for \grbs.
Woosely's ``failed Type Ib'' scenario involved accretion from a 
torus onto a central black hole and associated neutrino production
with mild collimation affecting the energetics and event rates.
Paczy\'nski's (1997) ``hypernova'' scenario is generically related by
invoking a ``failed Type Ib'' that makes a black hole, but invokes
a large magnetic field to yield a ``dirty fireball'' with 
relatively large energy and again mild collimation. With strong 
collimation, the energy could correspond to that of a supernova 
with proportional effect on the event rates (Shapiro 1998). We suggest
that it is successful SN Ib/c that make \grbs.

If \grbs, both at high and low redshift, are indeed produced by SN Ib/c, 
highly collimated, relativistic jets are required to solve the energy 
problem for the distant bursts. 
The hypothesis that all core collapse events produce 
relativistic beams is consistent with the current data.  We would 
expect that such beams are stopped by the massive envelopes of
SN II, but that strong asymmetries are generated.  This is
consistent with the observations that all SN II are polarized
and that almost no SN Ia are (Wang et al. 1996) and that SN II are
not associated with \grbs.  
For progenitors that have lost most or all of their
envelopes, the relativistic beams may survive to impact 
on the surroundings and create the phenomena associated
with \grbs. Thus all SN Ib/c may produce \grbs\ and all \grbs\ may be 
associated with SN Ib/c.

This interpretation requires that there are both isotropic and
non-isotropic components to the \grbs\ and associated phenomena. 
The isotropic component may be neither collimated nor Lorentz boosted
and beamed.  The anisotropic component must be subject to both.
The \gr\ energy associated with the SN Ib/c, $\sim10^{46}-10^{48}$ ergs, 
is extremely faint compared with the other BeppSAX 
\grbs\ for which we know something about their distances.  
It is also a minor fraction of the energy released 
during a supernova event.  If $10^{50}$ ergs of \gr\ energy
were collimated into a cone of $10^{-3}$ of 4 $\pi$ steradians and 
pointed towards the Earth, we would see an apparent isotropic \grb\  
energy of 10$^{53}$ ergs, or a fluence of about 10 ergs/cm$^2$ 
at the distance of SN 1998bw. 
This implies that we are seeing in GRB 980425 the isotropic 
component of the $\gamma$-ray flux. In some other direction 
it would have presumably been a factor
of $\sim 10^5$ brighter.  
Such an event would be visible at redshift greater than 3. 
Since only 1000 supernovae are known, it is very likely that a strongly 
collimated event would have escaped direct detection in a subset of some
types of supernovae. Likewise, these events would not be seen at 
cosmological distances unless
we are closely in the line-of-sight of the beam. 
The collimation required in this picture, confinement to an opening angle of
a few degrees, is not unprecedented.  Jets of comparable collimation
occur in many astronomical environments.

The ratio of the luminosities of the collimated, boosted 
component and the isotropic component may be related to the 
{\it apparent} energy ratios, assuming both components to be 
isotropic.  Comparing GRB 971214 (Kulkarni et al. 1998a) to 
GRB 980425 gives a ratio of $3\times10^{53}/7\times10^{47}$ or 
$4\times10^{5}$. If this is a representative ratio, then  
for the same detection limits the volume
sampled by the distant collimated, boosted component is about 
$3\times10^8$ times that of the nearby component.  
If the distant component is collimated into only $\sim10^{-3}$ 
of 4$\pi$ steradians, 
there should be about $3\times10^5$ distant, 
strongly collimated \grbs\ for every ``local"
isotropic event. This means that in the current BATSE catalog
there should be very few local events. 
The BATSE catalog should thus be 
dominated by the distant, collimated events, in complete accord
with the isotropy of that sample.
On the other hand, we apparently do see a few local events, 
the SN Ib/c of Table 1. To make our hypothesis plausible, the
luminosity ratio of the collimated, beamed component to the
isotropic component must be closer to $\sim10^4$.  There may
be a distribution of collimation and Lorentz factors and 
boosted luminosities and corresponding Malmquist bias so 
that GRB 971214 is not typical. 

From the brightness temperature 
Chevalier (1998) and Kulkarni et al. (1998b) have deduced
that the radio source associated with SN 1998bw is expanding 
relativistically. It is surprising that a supernova 
produced relativistic ejecta. 
If the radio flux is reduced by a factor of 
$\sim10^{4}$, one gets the characteristic radio flux
for SN Ib/c. The radio emission from SN 1998bw could be from the jet,
but with a viewing angle greater than zero, or perhaps after the jet has
slowed and the radio emission is more nearly isotropic
so there is less boost.  A component of the radio could arise 
from more normal radio emission associated with collision of 
supernova ejecta with the circumstellar medium. We note
that when it initially forms, the jet from the supernova core
might be unstable, yielding beaming in time-dependent directions
whereas the radio arises in a later blast wave phase when the
propagation and beaming directions are well established.
At the distance of SN 1998bw, radio interferometry should be able to 
resolve the jet and give some clues to its structure and anisotropy.

Spectropolarimetry of supernovae has already given evidence for
strong anisotropy of the core collapse process.  As remarked
above, all SN II and SN Ib/c are polarized, nearly all SN Ia
are not (Wang et al. 1996). There could be a number of reasons
why the ejecta of supernovae are asymmetric, but the hypothesis
that the core collapse process itself is a major contributor
yields the prediction that for smaller buffering envelopes,
the polarization should be greater.  Indeed, SN Ic 1997X produced
the largest polarization of any supernova yet measured, $\sim$
6 - 7 percent (Wang, Wheeler \& H\"oflich 1998).  

Core collapse supernovae may thus be highly anisotropic.
This may be seen in the observations of SN 1997ef and SN 1998bw.
From the width and temporal evolution of spectral features, 
SN 1997ef showed very high velocities at early times and
a spectrum unlike other identified supernovae (Garnavich et al. 1997).  
This may mean that it was observed more nearly on axis.  Polarization should
be maximized when an event is viewed normal to the elongation axis
and, indeed, SN 1997ef showed rather little polarization
(Wang, Howell \& Wheeler 1998).  SN 1997ef presumably
cannot have been beamed directly at us or we would have seen a very 
much brighter \gr\ source.  The hydrodynamics and the \gr\ beam
could plausibly have different opening angles.  We would predict
that SN 1998bw would also have relatively small polarization
since the radio emission and the optical spectra suggest it was
observed relatively pole-on.

If all core collapse events produce jets containing $\sim 10^{50}$ ergs
of energy, about $10^{-3}$ of the neutron star binding energy,
then current calculations of core collapse may need to be revised.
Core collapse must probably
include rotation and magnetic fields as intrinsic components.
The question of how to make a tightly collimated jet from
core collapse will clearly require separate effort.  We note
two relevant issues.  Pulsars show typical runaway velocities
of 500 \kms, requiring a kinetic energy of about $10^{49}$ erg.
Pulsar runaway might well be linked to the momentum imbalance
involved in generating two not quite equal jets with total
energy of about $10^{50}$ ergs.  The formation of jets might
also be enhanced if the magnetic field were larger than
often assumed.  Models for soft \gr\ repeaters require 
fields of $\sim 10^{15}$ Gauss (Duncan \& Thompson 1992).
Recent detection of a spin down in a soft \gr\ repeater
is consistent with such a large field (Kouveliotou et al. 1998).
The calculation of LeBlanc and Wilson (1970) of a rotating, 
magnetic core yielded a magnetic field of $\sim10^{15}$ gauss
and an axial jet of with an energy of $\sim10^{50}$ ergs.
Blackman \& Yi (1998; see also Thompson, 1994; M\'esz\'aros \& Rees 1997b) 
have noted that the energy in a jet associated
with a pulsar might be in the form of large amplitude
electromagnetic waves, rather than relativistic particles, {\it per se}.

By the hypothesis presented here, SN 1987A must have created an embryonic
jet in its core.  Even if this hypothesis is correct, it is not clear 
how such a jet could penetrate the hydrogen envelope, which is
known to have been massive.  Nevertheless, it might be 
appropriate to reexamine the ``mystery spot" in SN 1987A.

The authors are grateful to Peter Diener, Rob Duncan, and
Ed Robinson for helpful discussions and to the Crown \& Anchor for 
hospitality as the ideas presented here were developed. 
This research was supported in part by NSF Grant 95-28110,
a grant from the Texas Advanced Research Program, and by
NASA through grant HF-01085.01-96A from the Space Telescope
Science Institute which is operated by the Association of
Universities for Research in Astronomy, Inc., under NASA
contract NAS 5-26555.

\vfill\eject

\begin{table}

\caption{ Supernovae and GRB Association}

\scriptsize
\begin{tabular}{lrlllll}
\hline
\hline
{SN/GRB}    &{Type}   &  {Dates}&{RA (2000.0)}  
& {DEC (2000.0)} &{Radio} &{Prob.} \\ 
\hline
SN 1992H &  II     & 920204-930211 & 13 56 19.6   & +47 14 34.1& No  & {14.8}          \\
\hline
SN 1992ad&  Ib     & 920609-920605 & 12 26 49.6   & +08 52 38.7& Yes &                            \\
GRB 920609&        &               & 13 06 19.2   & -05 34 48.0&     & \ul{29.8,14.0},{1.7} \\
\hline
SN 1993J &  IIt    & 930323-930328 & 09 55 24.8   & +69 01 13.7& Yes &                            \\
GRB 930324&        &               & 12 00 48.0   & +59 34 12.0&     & \ul{7.7,3.5},{0.9}   \\
\hline
SN 1994D &  Ia     & 940328-940307 & 12 34 02.3   & +07 42 05.7& No  & {16.2}         \\
\hline
SN 1994I &  Ib/c   & 940329-940402 & 13 29 54.1   & +47 11 30.5& Yes &                            \\
GRB 940331&        &               & 10 31 14.4   & +57 31 48.0&     & \ul{8.1},{3.7}       \\
\hline
SN 1995D & Ia      & 950131-950109 & 09 40 54.8   & +05 08 26.2&     & {22.3}          \\
\hline
SN 1996B & IIt     & 951219-960109 & 12 23 56.6   & +48 46 42.0&     & {10.9}          \\
\hline
SN 1996N & Ib      & 960210-960224 & 03 38 55.4   & -26 20 04.0& Yes &                            \\
GRB 960221&        &               & 03 11 04.8   & -31 14 24.0&     & \ul{5.8,2.6},{0.7}   \\
\hline
SN 1996W & II      & 960402-960411 & 11 59 30.1   & -19 16 05.0& No  & {6.2}           \\
\hline 
SN 1996X & Ia      & 960315-960329 & 13 18 01.1   & -26 50 45.3& No  & {18.9}        \\
\hline
SN 1996cb& IIt     & 961208-961215 & 11 03 42.0   & +28 54 13.7& Yes & {2.4}         \\
\hline 
SN 1997ef& ?       & 971113-971125 & 07 57 02.8   & +49 33 40.2&     &                           \\
GRB 971115&        &               & 05 38 31.2   & +41 41 24.0&     & \ul{28.0,13.4},{3.5}\\
GRB 971120&        &               & 10 23 24.0   & +76 24 36.0&     & \ul{28.0},{13.4}    \\
\hline
SN 1997X& Ib/c     & 961231-970121 & 12 48 14.3   & -03 19 58.5& Yes &                           \\
GRB 970103&        &               & 11 56 19.2   & -06 51 00.0&     & \ul{13.8},{6.35}    \\
\hline
SN 1997ei& Ic      & 971103-971203 & 11 54 59.0   & +58 29 26.4&     &                            \\
GRB 971120&        &               & 10 23 24.0   & +76 24 36.0&     & \ul{36.1,17.8},{4.7} \\
\hline
SN 1997bp& Ia      & 970316-970401 & 12 46 53.8   & -11 38 33.2&     & {16.3}         \\
\hline
SN 1997bq& Ia      & 970401-970408 & 10 17 05.3   & +73 23 02.1&     & {21.1}          \\
\hline
SN 1997br& Ia      & 970320-970409 & 13 20 42.4   & -22 02 12.3&     & {46.8}         \\
\hline
SN 1997dd& IIt     & 970730-970813 & 16 05 46.0   & +21 29 14.0&     & {16.4}         \\
\hline
SN 1998S & IIn     & 980216-980302 & 11 46 06.0   & +47 29 00.0&     & {17.8}     \\
\hline
SN 1998T & Ib      & 980208-980303 & 11 28 33.2   & +58 33 43.7&     &                            \\
GRB 980218&        &               & 12 13 09.6   & +56 38 24.0&     & \ul{21.2},{10.0}    \\
GRB 980223&        &               & 14 45 09.6   & +28 12 36.0&     & \ul{21.2},{10.0}    \\
\hline
SN 1998bu& Ia      & 980428-980503 & 10 46 46.0   & +11 50 07.5&     & {0.197}   \\ 
\hline
\hline
\end{tabular}
\end{table}
\end{document}